\documentclass[aps,prx,reprint,superscriptaddress,nofootinbib]{revtex4-2}
\usepackage[T1]{fontenc}
\usepackage[latin9]{inputenc}
\usepackage{color}
\usepackage{babel}
\usepackage{amsmath}
\usepackage{amssymb}
\usepackage[unicode=true,pdfusetitle,
bookmarks=true,bookmarksnumbered=false,bookmarksopen=false,
 breaklinks=true,pdfborder={0 0 0},pdfborderstyle={},backref=false,colorlinks=true]
 {hyperref}
\usepackage{graphicx}
\hypersetup{citecolor=blue}
\usepackage{times}
\usepackage[breakable]{tcolorbox} 
\newtcolorbox{derivationbox}{
  breakable,
  boxrule=0.5pt,
  colframe=black!40,             % Hides the default box frame
  colback=white,
  left=3mm, right=3mm, top=2mm, bottom=2mm % Adjust padding
}
 
\begin{document}

\title{Nonequilibrium Theory for Adaptive Systems in Varying Environments}

\author{Ying-Jen Yang}
\email{ying-jen.yang@stonybrook.edu}
\affiliation{Laufer Center for Physical and Quantitative Biology, Stony Brook University, Stony Brook, NY 11794}

\author{Charles D. Kocher}
\altaffiliation{Now at Memorial Sloan Kettering Cancer Center, New York, NY 10065 
%\textcolor{red}{If we start a new submission somewhere other than PRX Life, I should probably change affiliation to MSK and add some acknowledgments.}\textcolor{blue}{Okay}
}
\affiliation{Department of Physics and Astronomy, Stony Brook University, Stony Brook, NY 11794}

\author{Ken A. Dill}
\affiliation{Laufer Center for Physical and Quantitative Biology, Stony Brook University, Stony Brook, NY 11794}
\affiliation{Department of Physics and Astronomy, Stony Brook University, Stony Brook, NY 11794}
\affiliation{Department of Chemistry, Stony Brook University, Stony Brook, NY 11794}

%\date{\today}

\begin{abstract}
Biological organisms are adaptive, able to function in unpredictably changing environments. Drawing on recent nonequilibrium physics, we show that in adaptation, fitness has two components parameterized by observable coordinates: a static \textit{Generalism} component characterized by state distributions, and a dynamic \textit{Tracking} component sustained by nonequilibrium fluxes. Our findings: (1) General Theory: We prove that tracking gain scales strictly with environmental variability and switching time-scales; near-static or fast-switching environments are not worth tracking. (2) Optimal Strategies: We explain optimal bet-hedging and phenotypic memory as the interplay between these components. (3) Control: We demonstrate, with an example, how to suppress pathogens by independently attacking their Generalism robustness (via environmental time fractions) and Tracking capabilities (via environmental switching speed). This work provides a physical framework for understanding and controlling adaptivity.
\end{abstract}

\maketitle

We seek design principles for the optimization and control of adaptive dynamics.  We consider the dynamics of a system, like a biological organism, that has some \textit{fitness} based on its ability to track dynamical change in its environment.  The ability to adapt to varying environments is fundamental for biological success \cite{murugan_roadmap_2021,levien_non-genetic_2021,bernhardt_life_2020}, prompting extensive theoretical \cite{kussell_phenotypic_2005,wolf_diversity_2005,donaldson-matasci_when_2013,belete_optimality_2015,xue_environment--phenotype_2019} and experimental \cite{acar_stochastic_2008,grantham_case_2016,gremer_within-and_2016,maxwell_when_2017} investigation into organisms' adaptive strategies. These strategies can be broadly classified into two types: (1) \textit{Tracking} strategies, where the organism senses the environment and shifts its phenotype or internal state according to the current \cite{lambert_memory_2014,nevozhay_mapping_2012} or anticipated \cite{saigusa_amoebae_2008,tagkopoulos_predictive_2008,mitchell_adaptive_2009,whitehead_diurnally_2009,marzen_optimized_2018,jabbur_bacteria_2024} conditions, and (2) non-tracking strategies (or ``\textit{Generalism},'' in the broad sense), including the ``pure'' generalist  (a single robust phenotype that does not switch) \cite{wang_evolving_2019,sachdeva_tuning_2020}  and stochastic bet-hedging  (random, environment-independent switching between phenotypes) \cite{slatkin_hedging_1974,venable_bet_2007,grimbergen_microbial_2015,muller_bet-hedging_2013,solopova_bet-hedging_2014}. Intuitively, Generalism is akin to wearing a versatile jacket suitable for most weather, while Tracking is the extra gain from actively changing outfits to match the specific environment.\\

Previous work has analyzed and computed optimal strategies for given environmental scenarios \cite{kussell_phenotypic_2005,mayer_diversity_2016,hufton_intrinsic_2016,hufton_phenotypic_2018,dinis_pareto-optimal_2022} and revealed (phase) transitions to distinct optimal states in different environmental settings \cite{mayer_transitions_2017,skanata_evolutionary_2016,patra_emergence_2015,xue_environment--phenotype_2019}. However, merely identifying the optimal strategy offers limited guidance on the underlying mechanisms and how a suboptimal organism can adjust to improve long-term growth. Knowing the location of a summit does not inherently reveal the trail to reach it. To address this gap, we need a navigational map that identifies the distinct physical components of the ``fitness elevation,'' thereby revealing axes for optimization and control. \\

In this work, we demonstrate that the long-term growth rate of a population %\textcolor{red}{population?} 
adapting to a varying environment can be decomposed into the exact sum of contributions from the two distinct strategies: 
(1) a static Generalism term, derived from the overlap between system ($\pi_x$) and environmental ($\pi_E$) state occupancies, and (2) a dynamic Tracking term, strictly proportional to the nonequilibrium probability flux $J$ scaled by the environmental switching time scale $\mathcal{T}_{\text{env}}$, \begin{equation} 
\text{Fitness} = \underbrace{\text{Generalism } (\pi_E ~\pi_x)}_{\text{Static~/~Equilibrium}} + \underbrace{\text{Tracking } (\mathcal{T}_{\text{env}}~ J).}_{\text{Dynamic~/~Nonequilibrium}} 
\label{eq: fitness parsing conceptual}
\end{equation}
Drawing on landscape-flux theory \cite{ao_potential_2004,wang_potential_2008,yang_potentials_2021} and the recent Caliber Force Theory  \cite{yang_principled_2025},
we identify $\pi_x$ and $J$ as the \textbf{observable coordinates} for adaptive dynamics. Just as latitude and longitude are needed to navigate a map, Generalism ($\pi_x$) and Tracking ($J$) are the orthogonal axes needed to navigate fitness. This coordinate geometry reveals their separate tunability and the internal degeneracy in fitness (Example 1). \\

The utility of this coordinate map lies in the distinct physical nature of its components. \textbf{First}, because Generalism is characterized by the static distribution $\pi_x$ and Tracking from the dynamic flux $J$, they exhibit different sensitivities to system parameters, allowing us to explain optimal adaptive strategies as the interplay of these two axes, such as bet-hedging \cite{kussell_phenotypic_2005,belete_optimality_2015} (Example 2) and phenotypic memory \cite{skanata_evolutionary_2016} (Example 3). 
\textbf{Second}, the two components are governed by different environmental statistics: Generalism depends on environmental bias ($\pi_E$), while Tracking is limited by the environmental switching time scale ($\mathcal{T}_{\text{env}}$). This separation suggests targeted control strategies, such as optimizing drug regimens to suppress pathogens by independently attacking their Generalism versus Tracking capabilities (Example 4). 
\textbf{Third}, this decomposition provides a practical shortcut for gauging the relative contributions of each strategy. 
Without the need to measure the fluxes ($J$), the Tracking contribution can be computed from the total growth rate by subtracting the measurable Generalism term. By mapping fitness onto these physical coordinates, we offer a framework to design adaptive strategies and inform the rational control of systems adapting to varying environments.

\section*{Theoretical Framework for Fitness Parsing}

We first quantify fitness. For an adaptive system reaching a statistical steady state under a varying environment,\footnote{The steady state has been called the ``red queen'' scenario \cite{van_valen_new_1973,ao_laws_2005,zhang_potential_2012,qian_fitness_2014}, a situation where a species constantly adapts to a changing environment.  This term comes from Alice in Wonderland; it is a race in which you must always keep running as fast as you can just to ``stay in place'' (stay alive).} fitness is naturally quantified as the Long-Term logarithmic Growth Rate of the population (LTGR). By time additivity and ergodicity, this growth rate is exactly equal to the long-term average of the per capita growth rate $g(E,x)$:
\begin{equation}
    \text{Fitness} = \lim_{t \to \infty} \frac{1}{t} \ln \frac{N(t)}{N(0)} = \sum_{E,x} \pi_{E,x}~ g(E,x)
\end{equation}
Here, $\pi_{E,x}$ represents the joint probability between the system's state $x$ and the environment's state $E$. The state variables $(E,x)$ encode the time dependency of $g$: $g[E(t),x(t)]=\dot{N}(t)/N(t)$.\\

\textbf{Environments with low variability are not worth tracking.}
We analyze how much of the statistical alignment $\pi_{E,x}$ (or \textit{synchrony}) is contributed from an independent baseline $\pi_x \pi_E$ (Generalism) and how much is from the additional gain from Tracking. By expressing the joint probability using marginal distributions ($\pi_E$ and $\pi_x$) and conditional distributions ($\pi_{x|E}=\pi_{E,x}/\pi_E$), a general parsing of the system-environment synchrony can be derived with a simple rewriting (Appendix A1 \cite{SI}):
\begin{equation} \label{eq: STR}
    \pi_{E,x}=\pi_E \, \pi_x+ \sum_{E' (\neq E)} \pi_E \, \pi_{E'} \, (\pi_{x|E}-\pi_{x|E'}).
\end{equation} 
The first term on the right represents a baseline that can be matched with a system independent of the environment (Generalism), whereas the second term represents how much a state is preferred in environment $E$ opposed to other environments $E'$ (Tracking). This latter term is the additional contribution from system-environment coupling.\\

Eq. \eqref{eq: STR} immediately offers an insight into environmental statistics: the Tracking term scales with the ``variability of the environment'' $\pi_E \pi_{E'}$. If the environment is highly biased toward one state (i.e., one $\pi_E\rightarrow1$ and others $\rightarrow 0$), the Tracking term diminishes. Thus, environments that are predominantly static (low variability) are not worth tracking. \\
%Below, we show that the tracking term---sustaining the nonzero conditional distribution differences ($\pi_{x|E} \neq \pi_{x|E'})$---requires nonequilibrium probability fluxes.

\textbf{Rapidly changing environments are not worth tracking.}
%\textbf{To track its environment, a system requires nonequilibrium fluxes.}
To relate Tracking to nonequilibrium flux,
we suppose that the environment-system pair follows a continuous-time Markov jump process. In each infinitesimal time step $\text{d}t$, only two types of transitions are possible: either the system changes $(E,x)\mapsto(E,x')$ with rates $R_{xx'}(E)$ or the environment changes $(E,x)\mapsto(E',x)$ with rates $R_{EE'}(x)$. A simultaneous change is higher order, $\mathcal{O}(\text{d}t^2)$, negligible in continuous time.  This is known as a \textit{bipartite} process \cite{hartich_stochastic_2014,horowitz_thermodynamics_2014,leighton_information_2024}. \\

{At the steady state, the total outward net flux from any joint state $(E,x)$ must be zero: \begin{align}
    0 =& \sum_{E'(\neq E)} \pi_{E,x} R_{EE'}(x) - \pi_{E',x}R_{E'E}(x)  \nonumber\\
    &+\sum_{x'(\neq x)} \pi_{E,x} R_{xx'}(E) -\pi_{E,x'}R_{x'x}(E). \label{eq: net flux balance}
\end{align}
For a given system state $x$, we assume that the \textit{environment evolves on its own} (i.e. $R_{EE'}(x)=R_{EE'}$ are independent of the system's state) and vectorize Eq. \eqref{eq: net flux balance} across all possible environmental states $E$: \begin{equation}
\mathbf{0}=-\boldsymbol{\pi}_{\text{joint}} ~\mathbf{R}_{\text{env}}+ \text{div}_{\text{sys}}~\mathbf{J} \label{eq: vectorize net flux balance}
\end{equation}
where $\boldsymbol{\pi}_{\text{joint}}=(\pi_{E_1,x},...,\pi_{E_N,x})$ is a row vector of the joint probability under different environments, $\mathbf{R}_{\text{env}}$ is the transition rate matrix of the environment, and $\text{div}_{\text{sys}}~\mathbf{J}$ is the row vector of system-wise net flux divergences (for each $E$, spreading out to all $x'(\neq x)$ from a fixed $x$). See Appendix A2 for details \cite{SI}. 
%\textcolor{red}{Writing out this divergence might be helpful. Similarly, from Eq 4, it seems to me that the signs in the R matrix are non-obvious. I think it would be helpful to have more details about how 4 goes to 5, probably in an appendix. I also think it would be useful for pi joint to have some indication that state x is fixed.} \textcolor{red}{UPDATE: I see that lots of this is defined in the Appendix. It might be useful to point to the appendix earlier and more clearly, so that people know where to look. Additionally, I would add something to the general derivation appendix section on how this divergence of J turns into just a regular J in our formulas.} \textcolor{blue}{Done.} 
Since $\boldsymbol{\pi}_{\text{env}}=(\pi_{E_1},...,\pi_{E_N})$ is the steady-state distribution of the environment, satisfying $\boldsymbol{\pi}_{\text{env}}\mathbf{R}_{\text{env}}=\mathbf{0}$, we can expand $\boldsymbol{\pi}_{\text{joint}}$ around the decoupled baseline $\boldsymbol{\pi}_{\text{env}}\pi_x$. ``Inverting'' $\mathbf{R}_{\text{env}}$  yields: \begin{equation}
    \boldsymbol{\pi}_{\text{joint}}=\boldsymbol{\pi}_{\text{env}} ~\pi_x+(\text{div}_{\text{sys}}~\mathbf{J}) (\tilde{\mathbf{R}}_{\text{env}}^{-1}) \label{tracking is prop to J/R}
\end{equation} where $\tilde{\mathbf{R}}_{\text{env}}^{-1}$ is the Drazin inverse (or generalized inverse) of the environmental transition rate matrix $\mathbf{R}_{\text{env}}$. This inverse matrix captures the relaxation time of the environment. 
It is constructed via spectral decomposition $\tilde{\mathbf{R}}_{\text{env}}^{-1}=\sum_{\lambda_k \neq 0} \lambda_k^{-1}  \mathbf{v}^{\mathsf{T}}_k \mathbf{u}_k$ where  $\lambda_k$ are the eigenvalues of $\mathbf{R}_{\text{env}}$ with corresponding left ($\mathbf{u}_k$) and right ($\mathbf{v}_k^{\mathsf{T}}$) eigenvectors.  The reciprocals of these eigenvalues are proportional to the environmental time scales $\mathcal{T}_{\text{env}}$. Thus, under fast environmental switching ($\mathcal{T}_{\text{env}} \to 0$), the Tracking contribution vanishes. }\\

This result mirrors the equilibrium-nonequilibrium splitting of dynamics in stochastic thermodynamics \cite{seifert_entropy_2005, esposito_three_2010,ge_physical_2010}. By demonstrating that Tracking necessarily involves nonequilibrium fluxes, our parsing clarifies the fundamental connection between time irreversibility and adaptive dynamics \cite{mustonen_fitness_2010,zhang_potential_2012,qian_fitness_2014,kobayashi_fluctuation_2015,genthon_fluctuation_2020, rao_evolutionary_2022}.
However, viewed through the lens of landscape-flux theory \cite{ao_potential_2004,wang_potential_2008,yang_potentials_2021} and the recent Caliber Force Theory \cite{yang_principled_2025}, Eq. \eqref{tracking is prop to J/R} reveals a deeper structure: Generalism ($\pi_x$) and Tracking ($J$) constitute thermodynamics-like \textbf{observable coordinates} for adaptive dynamics and are distinct degrees of freedom that can be \textbf{tuned independently}. Moreover, this coordinate mapping highlights a \textbf{design degeneracy} in adaptive dynamics allowing bidirectional transitions: since fitness is fully determined by Generalism $\pi$ and Tracking $J$, the remaining degrees of freedom---the noise strength in diffusion \cite{ao_potential_2004,wang_potential_2008,yang_potentials_2021} or the ``traffic'' (sum of both way fluxes) in Markov jumps \cite{maes_frenesy_2020,yang_principled_2025}---can be varied without altering fitness, provided that $(\pi_x,J)$ remain constant (Example 1). \\

%\textbf{Summary:} We have derived that the long-term synchrony of general Markov adapting systems parses into a static generalism baseline and a dynamic tracking gain. While Generalism depends only on stationary statistics ($\pi_E, \pi_x$), Tracking is governed by the interplay of environmental variability and relative switching speeds (Eq. \ref{eq: STR} and \ref{tracking is prop to J/R}). This decomposition maps fitness onto two controllable axes, enabling the targeted optimization and control strategies we explore in the following examples.

\section*{Example 1: Adapting to a 2-State Markov Switching Environment} 

\begin{figure}
\begin{centering}
\includegraphics[width=\columnwidth]{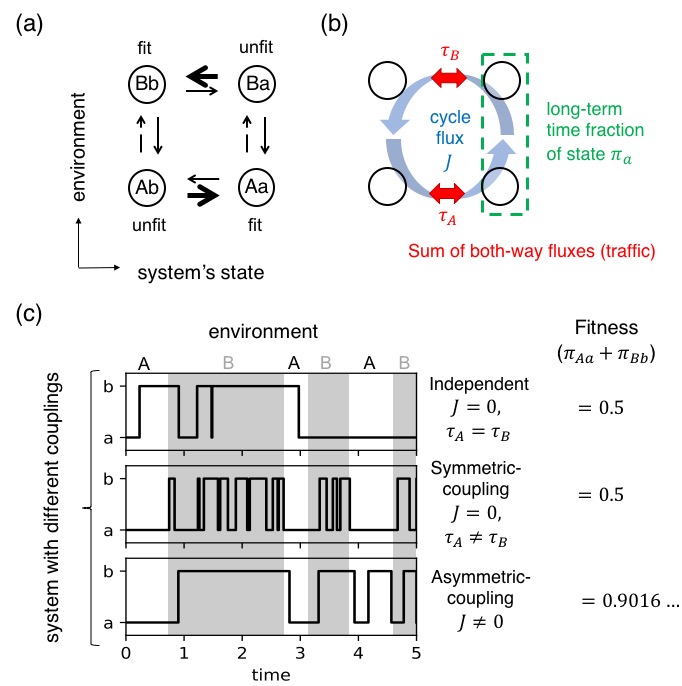}
\par\end{centering}
\caption{\textbf{The state space, observables, and types of system-environment coupling of a simple adapting individual.} \textbf{(a)} The total state space of the two-state system $\{a,b\}$ under a Markov environmental switch $\{A,B\}$. Vertical arrows are environmental transitions and horizontal arrows are system transitions. \textbf{(b)} Four statistical observables that fully parameterize the system's dynamics. $\pi_a$ is the long-term time fraction that the system is in $a$ regardless of the environmental state. Traffic $\tau_E$ are the sum of fluxes in both direction under environment $E$. $J$ is the net cycle flux. \textbf{(c)} Three types of system-environmental coupling classified by the observables. Net flux $J$ is needed to track the environment and outperform an independent switch in the synchronization task. {Fitness values shown on the right are the long-term time fraction of being at the synchronized (fit) states. Parameters used in (c) are as follows: $R_{AB}=R_{BA}=1$, $g_a(A)=g_b(B)=1$ and $g_a(B)=g_b(A)=0$ for all three cases. $R_{ab}(E)=R_{ba}(E)=1, E=A \text{ or } B,$ for the independent case; $R_{ab}(A)=R_{ba}(A)=0.1,R_{ab}(B)=R_{ba}(B)=10$ for the symmetric-coupling case; $R_{ab}(A)=0.2,R_{ba}(A)=10,R_{ab}(B)=10,R_{ba}(B)=0.2$ for the asymmetric-coupling case.  }\label{fig: square}}
\end{figure}

Here, we illustrate our framework with a minimal model. Suppose an individual in the population can take either of two phenotypic states $\phi \in \{a,b\}$ under a two-state switching environment $E \in \{A,B\}$. State $a$ is more fit in environment $A$, and state $b$ is more fit in $B$. For a Markov switching environment and system, the joint state space is a square with four nodes and four edges (Fig. \ref{fig: square}a). This model can describe, for example, a single chemo-sensing receptor in \textit{E. coli} \cite{barato_efficiency_2014}, or---our focus here---phenotype-switching individuals in a population under the slow growth limit.\\

\textbf{Slow Growth Approximation.}
Strictly speaking, the state of an adaptive population is the vector $\mathbf{x}=(N_a, N_b)$, which determines the instantaneous growth rate of the total population $N(t)=N_a(t)+N_b(t)$. However, the population composition is dynamically skewed by the interplay of switching and differential growth (i.e., fitter phenotypes accumulate faster and skew the distribution). While we will address this full population dynamics in later examples, here we simplify the problem by focusing on the individual phenotype $\phi \in \{a,b\}$ and the phenotype switching. Let $g_\phi$ denote the growth rate and $R_{\phi \phi'}$ the switching rates. We consider the limit where individual replication is much slower than switching ($g_\phi \ll R_{\phi \phi'}$). Under this time-scale separation, the relevant state description reduces from the population counts $\mathbf{x}$ to the individual phenotype $\phi$, as the population composition effectively relaxes to the quasi-static switching distribution $\pi_{E,\phi}$ before differential growth can significantly distort it. This allows us to approximate the population fitness simply as the expected replication rate of a single stochastic individual in this example:
\begin{equation} \label{eq: approxi LTGR in slow growth}
\text{LTGR} = \lim_{t\rightarrow \infty} \frac{1}{t} \ln \frac{N(t)}{N(0)} \approx \sum_{E,\phi} \pi_{E,\phi} ~g_{\phi}(E).
\end{equation}
This approximation effectively transforms the problem from solving coupled population differential equations into a simpler statistical task of computing averages over the single-particle stationary distribution.\\

\textbf{Fitness Decomposition.}
We now apply the general decomposition derived in the Theory setion (Eqs. \ref{eq: STR} and \ref{tracking is prop to J/R}) to this system. The Generalism term can be obtained by simply replacing $\pi_{E,\phi}$ with $\pi_E \pi_\phi$ in Eq. \eqref{eq: approxi LTGR in slow growth}, leading to $\pi_a \bar{g}_a + \pi_b \bar{g}_b$ where $\bar{g}_\phi = \pi_A g_{\phi}(A)+\pi_B g_{\phi}(B)$ is the environment-averaged growth rate. The Tracking term requires the Drazin inverse of the environmental transition matrix. For a 2-state environment, the transition matrix has only one non-zero eigenvalue $\lambda = -(R_{AB} + R_{BA})$, and the Drazin inverse is simply the inverse of this relaxation rate multiplied by the outer product of its eigenvectors.  With details shown in  Appendix B1 \cite{SI}, combining Eqs. \eqref{tracking is prop to J/R} and \eqref{eq: approxi LTGR in slow growth} yields:
\begin{equation} \label{eq: LTGR decomposed in pi-J}
\text{LTGR} = \underset{\text{Generalism}}{\underbrace{\pi_a~\bar{g}_a+ \pi_b~ \bar{g}_b}} + \underset{\text{Tracking}}{\underbrace{\left[ \pi_A \pi_B~ \bar{T} \right] J ~\Delta_g}}
\end{equation}
where $\bar{T} = R_{AB}^{-1} + R_{BA}^{-1}$ is the average environmental cycle period and $\Delta_g = [g_{a}(A)-g_{b}(A)]+[g_{b}(B)-g_{a}(B)]$ is the growth advantage of being in the correct state.
This confirms our general theory: Tracking gain is proportional to the product of Environmental Variability ($\pi_A \pi_B$), Environmental Time Scale ($\bar{T}$), and the system's Nonequilibrium Flux ($J$).\\

\textbf{Fitness Degeneracy.}
While Eq. \eqref{eq: LTGR decomposed in pi-J} shows that fitness is determined entirely by $\pi_a$ and $J$ (given a fixed environment), these two coordinates are insufficient to fully describe adaptive dynamics. The system has four independent transition rates ($R_{ab}(A), R_{ba}(A), R_{ab}(B), R_{ba}(B)$), and thus requires four statistical observables for a complete parameterization. As illustrated in Fig. \ref{fig: square}(b), the complete set are:
\begin{enumerate}
\item The total occupancy $\pi_a$ (Generalism coordinate).
\item The net cycle flux $J$ (Tracking coordinate).
\item Two ``traffic'' terms $\tau_A$ and $\tau_B$, representing the sum of bidirectional fluxes in each environment, $\tau_E = \pi_{E,a}R_{ab}(E) + \pi_{E,b}R_{ba}(E)$ \cite{maes_frenesy_2020,yang_principled_2025}.
\end{enumerate}
These observable coordinates parameterize the transition rates $R_{xx'}(E) = (\tau_E \pm J)/(2\pi_{Ex})$, and this dimension accounting reveals a fundamental {fitness degeneracy}. The ``traffic'' difference $|\tau_A - \tau_B|$ corresponds to a symmetric coupling with the environment (e.g., the system switches faster in $A$ than in $B$ in \textit{both} directions). While such symmetric coupling indicates that the system ``senses'' the environment, Eq. \eqref{eq: LTGR decomposed in pi-J} shows it contributes nothing to fitness. Only the asymmetric, nonequilibrium coupling---manifested as non-zero flux $J$---enables the system to track and outperform an independent generalist (Fig. \ref{fig: square}c). \\

The traffic terms $\tau_E$ serve only as kinetic constraints, limiting the maximum possible Tracking flux but not determining the direction of adaptation itself. This is because the amplitude of net flux on each edge $|J_{ij}|=|p_{ij}-p_{ji}|$---where $p_{ij}$ is the one way flux from $i$ to $j$---can not surpass the traffic $\tau_{ij}=p_{ij}+p_{ji}$. This dynamical efficiency is exactly what the steady-state ``entropy production'' in stochastic thermodynamics \cite{ge_physical_2010,esposito_three_2010, seifert_entropy_2005}, $\sum_{i<j} J_{ij} ~\ln[(\tau_{ij} +J_{ij})/(\tau_{ij}-J_{ij})]$ measures. See \cite{yang_principled_2025} for more discussions. \\

This example illustrates, in the simplest setting, the two implications of mapping fitness onto observable coordinates: separate tunability (Generalism $\pi$ and Tracking $J$) and fitness degeneracy (traffic $\tau$).

\section*{Example 2: Achieving Optimal Bet-Hedging}

\begin{figure}
\begin{centering}
\includegraphics[width=0.8 \columnwidth]{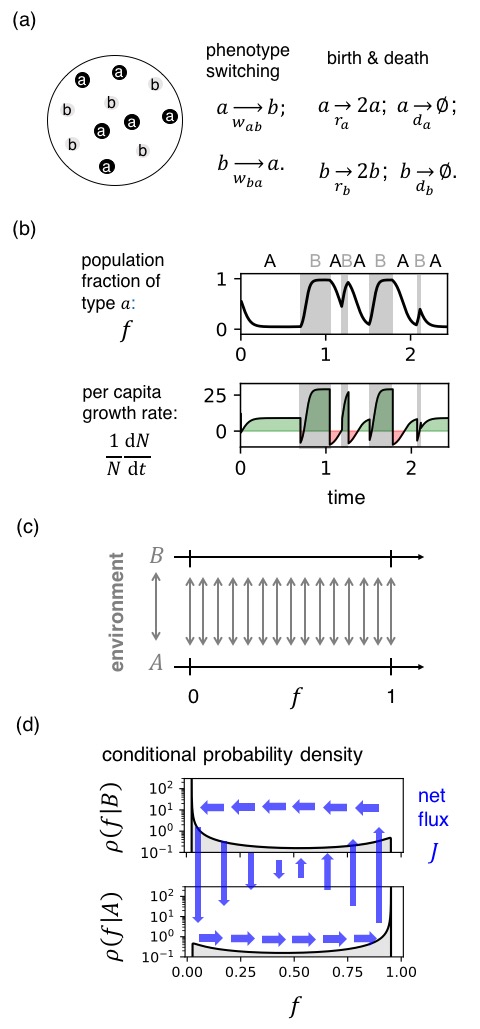}
\par\end{centering}
\caption{\textbf{The dynamics and total state space for a phenotype-switching population.} \textbf{(a)} We consider the population dynamics of two-state individuals with Poisson birth-death and switching. \textbf{(b)} The population type fraction (of $a$) tracks the environmental changes and determines the instantaneous per capita growth rate. \textbf{(c)} The system-environment total state space consists of two line segments $f\in [0,1]$, one for each environment $A$ and $B$. The population can jump between the two segments at any $f$. \textbf{(d)} The (cyclic) net flux over the total state space leads to the long-term conditional probability density $\rho(f|E)$, forming a flux-tracking relation for population dynamics.
{Parameters used in (b) and (d) are: $R_{AB}=2$, $R_{BA}=1$, $g_a(A)=30$, $g_b(B)=10$, $g_a(B)=g_b(A)=-10$, and $w_{ab}(E)=w_{ba}(E)=1$.}
\label{fig: evo example}}
\end{figure}

We now go beyond the slow-growth limit in Example 1 to consider a large phenotype-switching population where individuals can reproduce, die, or switch phenotype under a Markov switching environment, as illustrated in Fig. \ref{fig: evo example}(a). Commonly used to model phenotypic switching in microbes like \textit{E. coli} \cite{kussell_phenotypic_2005,skanata_evolutionary_2016} or yeast \cite{belete_optimality_2015} (or species reproducing with mutation; see Appendix C2 \cite{SI}), these dynamics are described by the following ODEs: 
\begin{subequations} \label{eqs: ODEs population}
\begin{align}
\frac{\text{d}N_{a}}{\text{d}t} & =g_{a} (E)N_{a}-w_{ab}(E) N_{a}+w_{ba}(E) N_{b} \; , \\
\frac{\text{d}N_{b}}{\text{d}t} & =g_{b}(E)N_{b}+w_{ab}(E) N_{a}-w_{ba}(E) N_{b}, \; 
\end{align}
\end{subequations}
where $g_x=r_x-d_x$ is the net growth rate of type $x$ with reproduction rate $r_x$ and death rate $d_x$, which we presume to be constant and independent of the population size. In this deterministic limit, the adaptive dynamics flow without diffusive backtracking. Thus, unlike the discrete Example 1, the additional ``traffic'' variables are not needed (they equal to net fluxes for one-way flows), and the system exhibits no fitness degeneracy. We note, however, that the traffic variables (and the associated degeneracy) would reappear if one analyzes the Poissonian stochastic dynamics before the large population limit---a level of detail of future interests but not required for the results derived here.\\

To extract the physical drivers of adaptation, it is standard to rewrite these equations in terms of more informative variables. Denoting the total population $N=N_a+N_b$ and the population fraction of the $a$ phenotype $f = N_a/N$, these ODEs become
\begin{subequations} \label{eqs: ODEs population in N-f}
\begin{align}
\frac{\text{d} f}{\text{d}t} & = f(1-f)~\Delta_{ab}  +(f_w-f) \sigma_w \label{eq: ODE f}\; , \\
\frac{\text{d}N}{\text{d}t} & = [g_a f + g_b (1-f)]~N \label{eq: ODE N}.\;
\end{align}
\end{subequations}
where $\sigma_w = w_{ab}+w_{ba}$ is the sum of the switching rates, $f_w = w_{ba}/\sigma_w$ is the steady-state phenotype fraction if there's no growth (or death), and $\Delta_{ab} = g_a - g_b$ is the growth rate difference, also known as the selection coefficient. All these parameters depend on the environment.
The rewriting shows that switching drives the type fraction $f$ toward $f_w$ with rate $\sigma_w$ whereas natural selection, via the other term of $\text{d}f/\text{d}t$,
%which takes the usual form \cite{kimura1969natural}, 
aims to fixate $f$ toward 0 or 1 with a rate $|\Delta_{ab}|$. The overall dynamics is a balance of the two. Most importantly, it shows that the type fraction $f$ is the sensing dimension that tracks the environmental changes, whereas the $N$ dynamics in Eq. \eqref{eq: ODE N} converts $f$ into the per capita growth rate as the population's average fitness. See Fig. \ref{fig: evo example}(b) for an example trajectory. \\

%\textbf{An optimization principle for stochastic bet-hedging.} 
\textbf{Fitness parsing for adaptive population:}
We now derive the fitness parsing for this deterministic adaptive system. First, observing that the per capita growth rate of $N$ in Eq. \eqref{eq: ODE N} is linear in $f$, the Long-Term Growth Rate (LTGR) is simply the time-average of the per capita growth rate. This average can be expressed as a summation of conditional averages  (Appendix A4 of SM \cite{SI}):
\begin{subequations}
\begin{align}
&\text{LTGR} = \lim_{t\rightarrow \infty} \frac{1}{t} \ln \frac{N(t)}{N(0)}\\
&=\sum_{\mathcal{E}} \pi_\mathcal{E} \left[g_{a}(\mathcal{E}) \langle f|\mathcal{E} \rangle + g_{b}(\mathcal{E}) (1-\langle f|\mathcal{E} \rangle)\right] \label{eq: LTGR in terms of conditional average f}
\end{align}
\end{subequations}
where $\pi_\mathcal{E}$ is the long-term occupancy of environment $\mathcal{E}$ and
\begin{equation}
\langle f|\mathcal{E} \rangle = \lim_{t\rightarrow \infty} \frac{\int_0^t ~f(s) ~\delta_{E(s),\mathcal{E}} ~\text{d}s}{\text{total time in }\mathcal{E}}
\end{equation}
is the conditional long-term empirical average of $f$ under environment $\mathcal{E}$.
The term $\delta_{E(s),\mathcal{E}}$ within the integral is a Kronecker delta function, which collapses the integral so that it is only taken over times when the system is in environment $\mathcal{E}$. For Markov switching environments, the conditional average $\langle f |\mathcal{E}\rangle$ has an analytical expression that can be computed numerically without ensemble simulations \cite{hufton_intrinsic_2016,hufton_phenotypic_2018,dinis_pareto-optimal_2022}, which we've used throughout the paper.\\

Algebraically rearranging Eq. \eqref{eq: LTGR in terms of conditional average f} allows us to separate the environment-averaged contribution (Generalism) from the environment-dependent difference (Tracking):
\begin{align}
\text{LTGR} =& \underbrace{\langle f \rangle \bar{g}_a + \langle 1-f\rangle \bar{g}_b}_{\text{Generalism}} + \underbrace{\pi_A \pi_B \left[ \langle f|A \rangle - \langle f|B \rangle\right] \Delta_g}_{\text{Tracking}} \nonumber %\label{eq: STR for LTGR}.
\end{align}
Our general theory connects the difference in conditional occupancy to nonequilibrium flux. Let $J(f,A)$ denote the probability current density (horizontal arrows in Fig. \ref{fig: evo example}(d)) and $\bar{J}=\int_0^1 J(f,A) \text{d}f$ be the total net flux. Then, we derive in Appendix B2 \cite{SI} that
\begin{equation} \label{SFR for LTGR in random environment}
\text{LTGR} = { \langle f \rangle ~\bar{g}_a+ \langle 1-f \rangle ~\bar{g}_b} + \pi_A \pi_B~ (\bar{T} \cdot  \bar{J})  ~\Delta_g.
\end{equation}
Fitness of a population is also spanned by the orthogonal coordinates of Generalism (occupancy $\langle f \rangle$) and Tracking (flux $\bar{J}$).\\

\textbf{Explaining Optimal Bet Hedging:}
Eq. \eqref{SFR for LTGR in random environment} generalizes the single-individual limit to the population level with temporal averages ($\pi_a, J$) replaced by spatiotemporal averages ($\langle f \rangle, \bar{J}$). Here, the cycle flux $\bar{J}$ represents the statistical circulation of the population composition driven by environmental changes.
Stochastic bet-hedging---the strategy of maintaining phenotypic diversity to avoid extinction  \cite{levien_non-genetic_2021}---enables this circulation to persist. When implemented via \textit{stochastic switching} rates ($w_{ab}, w_{ba}$) independent of the environment, this strategy prevents natural selection from fixing the population at $f=0$ or $1$, thereby generating a continuous Tracking advantage. \\

\begin{figure}
\begin{centering}
\includegraphics[width=\columnwidth]{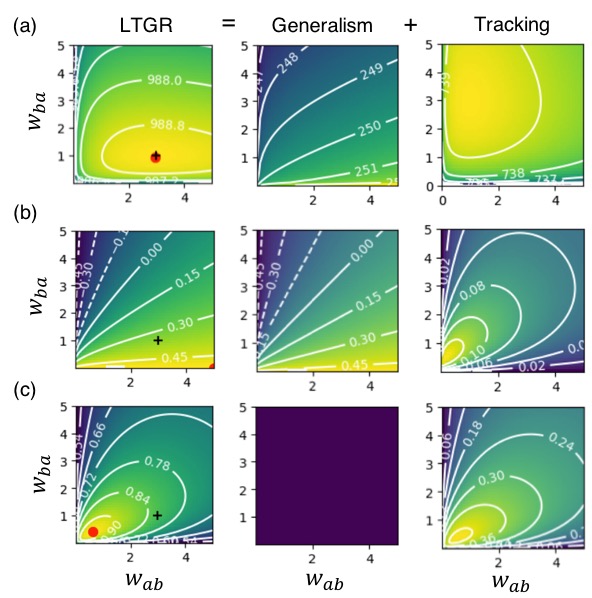}
\par\end{centering}
\caption{\textbf{Fitness Parsing explains the existence of optimal nonzero stochastic switching rates}. Analyzing the dependence of Long-Term (logarithmic) Growth Rate (LTGR) on the stochastic switching rates $(w_{ab},w_{ba})$. Black +'' denotes the position where $(w_{ab},w_{ba})=(R_{AB},R_{BA})$ whereas red o'' denotes the optimal $(w_{ab},w_{ba})$. \textbf{(a) Fast growth limit.}
The optimal switching rates match the environment's switching rates in this limit.
\textbf{(b) When the Generalism strategy is optimal.} The case where the growth rate is not fast.  The optimum LTGR happens at the boundary, with $w_{ba}=0$. \textbf{(c) Pure Tracking / Distinct Tunability.} This is a case where the environmental average growth rates of the two types are the same: $\bar{g}_a=\bar{g}_b=0.5$. The Generalism term becomes a flat surface, leaving the fitness landscape to be determined solely by the Tracking term.
\label{fig: SR evolution} Parameters used are the following: (a) $R_{AB}=3$, $R_{BA}=1$, $g_a(A)=1000$, $g_b(A)=-1000$, $g_a(B)=-1000$, $g_b(B)=1000$; (b) $g_a(A)=1$, $g_b(A)=-1$, $g_a(B)=-1$, $g_b(B)=1$. All other parameters are the same as (a); (c) $R_{AB}=3$, $R_{BA}=1$, $g_a(A)=5$, $g_b(A)=-1$, $g_a(B)=-1$, $g_b(B)=1.$ Parameter ranges are selected to illustrate the fast-growth limit (a) and the topological features of the fitness landscape (b,c).}
\end{figure}

This decomposition reconciles previous findings with a unified view. In the fast-growth limit, Kussell and Leibler showed that optimal switching rates match the environmental rates \cite{kussell_phenotypic_2005}. Our framework recovers this result (Fig. \ref{fig: SR evolution}a) and explains the parameter-dependent deviations observed in general regimes \cite{belete_optimality_2015}: the optimum is strictly determined by the balance between the Generalism and Tracking coordinates.\\

First, note that the Tracking term (proportional to $\bar{J}$) always forms a bell-shaped landscape in the space of stochastic switching rates (the right column in Fig. \ref{fig: SR evolution}). This shape arises from two competing physical limits. On one hand, flux vanishes when switching rates are tiny, as the selection force ($\text{d}f/\text{d}t \approx f(1-f)\Delta_{ab}$) drives the population to fixation. On the other hand, flux also vanishes when switching rates are extremely large, as rapid switching ($\text{d}f/\text{d}t \approx (f_w-f)\sigma_w$) screens out the environment-dependent selection required to drive the cycle. Consequently, the Tracking advantage is maximized only at intermediate rates.\\

Second however, note that the total fitness is not determined by the Tracking flux alone. The Generalism term $\langle f \rangle$ creates a competing landscape. This can mask the Tracking bell shape, pushing the optimum to the boundaries (pure Generalism), as seen in Fig. \ref{fig: SR evolution}(b). When the two environment-averaged growth rates are equal ($\bar{g}_a=\bar{g}_b$), the Generalism term becomes a flat surface (a constant $\bar{g}_b$). In this regime, changes in LTGR are driven {solely} by the Tracking coordinate. As shown in Fig. \ref{fig: SR evolution}(c), the emergence of an optimal nonzero switching rate here is pure flux maximization, unconfounded by Generalism.\\

This example demonstrates the fitness parsing in a population model, explaining optimal bet-hedging as the interplay between Generalism and Tracking.

\section*{Example 3: Achieving Optimal Phenotypic Memory.}

Some systems can \textit{learn and remember} something about their environments. Some microorganisms can sense their environment and switch their phenotypes accordingly, called \textit{responsive switching} \cite{kussell_phenotypic_2005}. {This responsive switching strategy combines the Tracking mechanisms seen in the previous two examples: the cyclic Tracking fluxes now arise from both responsive phenotype switching and differential growth.} For example, \textit{E. coli} can activate the \textit{lac} operon to metabolize lactose when it is present, and the bacteria can turn it off when more preferred nutrients like glucose are available. Using a simple phenomenological population ODE model, Skanata and Kussell showed that there is an optimal responsive switching rate (whose inverse is interpreted as the strength of phenotypic memory) under a stochastic switching environment \cite{skanata_evolutionary_2016}. Here, we show that fitness parsing can explain the mechanism behind the optimality and give the conditions for its existence.\\

Suppose $A$ is the lactose environment, $B$ is the glucose environment, $a$ is the phenotype expressing \textit{lac} proteins, and $b$ is the phenotype that does not. Then, following \cite{skanata_evolutionary_2016} and evaluating Eq. \eqref{eq: ODE f} for the two environments,  the equations for the fraction $f$ of type $a$ under $A$ and $B$ are\footnote{{This phenomenological model from Skanata and Kussell assumed that the phenotype switching scheme changes perfectly with the environment, with no delay in the switching strategy changes---from $(1-f)~\tau_{\text{on}}^{-1}$ to $(-f)~\tau_{\text{mem}}^{-1}$---and no back-stepping. It could be interesting to investigate how the optimality changes without these assumptions{, which is beyond the scope of the current work}. %\textcolor{red}{``but that analysis is beyond the scope of the current work''}
}}
\begin{subequations} \label{eqs: f ODE under lactose and glucose}
\begin{align}
    \frac{\text{d}f}{\text{d}t} &= f(1-f) h+(1-f)~\tau_{\text{on}}^{-1} ,\text{ under } A; \\
    \frac{\text{d}f}{\text{d}t} &= f(1-f) (-c) -f~\tau_{\text{mem}}^{-1} ,\text{ under }B
\end{align}
\end{subequations}
where $\tau_{\text{on}}$ is the time scale to sense and reach high-level \textit{lac} proteins,  $\tau_{\text{mem}}$ is the time scale for the degradation of \textit{lac} proteins (phenotypic memory), and the growth rates for the LTGR computation are $g_a(A)=h$, $g_b(A)=0$, $g_a(B)=g-c$, $g_b(B)=g$, with $g$ being the growth rate under glucose, $h$ being the growth rate of individuals with high \textit{lac} proteins in the lactose environment, and $c$ being the fitness cost for having high \textit{lac} proteins under the glucose environment. Of interest here is how tuning the memory level $\tau_{\text{mem}}$ affects the LTGR.\\

\begin{figure}
\begin{centering}
\includegraphics[width=\columnwidth]{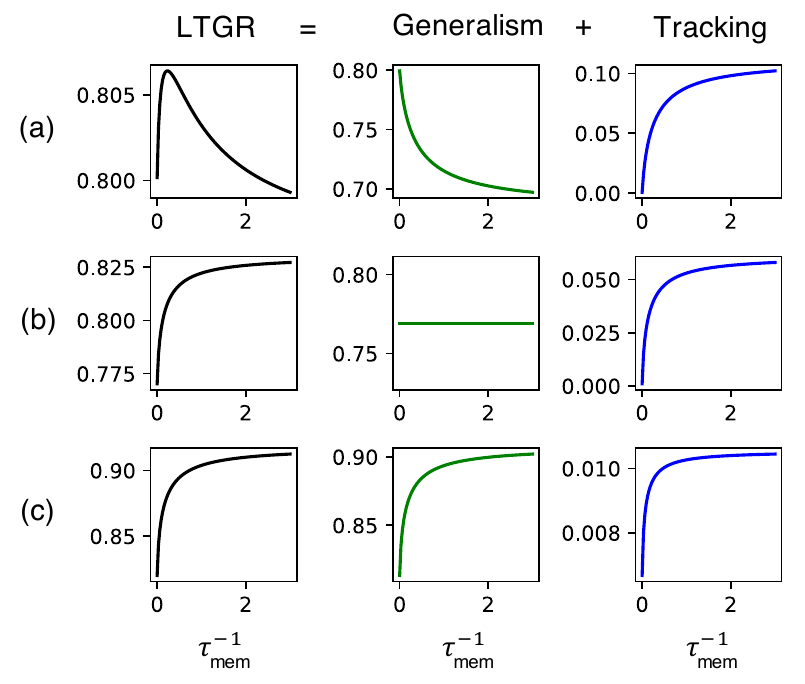}
\par\end{centering}
\caption{\textbf{Fitness Parsing explains the existence of optimal phenotypic memory.} From (a) to (c), we increase $R_{AB}$ from $2$ to $10/3$ to $10$. The average growth rates of the two environments change from $(\bar{g}_a,\bar{g}_b)=(0.8,0.\overline{6})$ to $\approx(0.769,0.769)$ and to $=(0.\overline{72},0.\overline{90})$. Optimal {at intermediate} phenotype memory 
cannot happen in (b) or (c) when $\bar{g}_a\le \bar{g}_b$. Note that the y-axes for the components are individually scaled to visualize the qualitative trends in each regime. Other parameters are chosen according to the phenomenological model in \cite{skanata_evolutionary_2016}: fixed at $R_{BA}=1$, $h=1$, {$\tau_{\text{ON}}^{-1}=1$}, $g=1$, $c=0.3$.
 All parameter values are chosen relative to the growth parameter $g$. \label{fig: optimal memory}}
\end{figure}

\textbf{Explaining the optimal phenotypic memory observed in \cite{skanata_evolutionary_2016}}: We show that a weaker memory (larger $\tau_{\text{mem}}^{-1}$) always increases the Tracking term by allowing the population to respond faster (the right hand side of Fig. \ref{fig: optimal memory}). However, it also lowers the average fraction $\langle f \rangle$ of lac-expressing cells. The Generalism term in the fitness parsing, $\langle f\rangle \bar{g}_a +\langle1-f\rangle \bar{g}_b$, dictates whether this decrease in $\langle f \rangle$ is beneficial or detrimental. Crucially, this implies that an optimal intermediate memory can only exist if maintaining the memory phenotype is generally beneficial ($\bar{g}_a > \bar{g}_b$, Fig. \ref{fig: optimal memory}a). In this regime, reducing memory hurts Generalism, creating the trade-off against the Tracking gain necessary for an intermediate optimal memory. Conversely, if $\bar{g}_a \le \bar{g}_b$ (Fig. \ref{fig: optimal memory}b-c), reducing memory is either neutral or beneficial for Generalism. With both components favoring faster switching, the trade-off vanishes, and no intermediate optimum exists.\\

%Eqs. \eqref{eqs: f ODE under lactose and glucose} indicate how tuning $\tau_{\text{mem}}$ affects the generalism term and the tracking term in the fitness parsing Eq. \eqref{SFR for LTGR in random environment}. A weaker memory (smaller $\tau_{\text{mem}}$)  means a faster switching rate $\tau^{-1}_{\text{mem}}$ from $a \mapsto b$ under the glucose environment $B$, leading to a smaller $\langle f \rangle$ and a bigger net flux. Thus, the flux contribution to LTGR will increase with increasing  rate $\tau_{\text{mem}}^{-1}$ as shown on the right hand side of Fig. \ref{fig: optimal memory}. Whether the decay of $\langle f \rangle$ increases or decreases LTGR depends on whether $\bar{g}_a-\bar{g}_b$ is negative or positive, as shown in the middle column of Fig. \ref{fig: optimal memory}. On the one hand, if $\bar{g}_a>\bar{g}_b$, the decay of $\langle f \rangle$ leads to a decrease in its contribution to the LTGR, and an intermediate maximum in fitness is then possible as shown in Fig. \ref{fig: optimal memory}(a). On the other hand, if $\bar{g}_a\le\bar{g}_b$---\textit{i.e.} if the environment-averaged growth rate of having many \textit{lac} proteins is not bigger than having none, as is the case in Fig. \ref{fig: optimal memory} (b-c), the contribution of $\langle f \rangle$ is non-decaying, and less memory (larger $\tau_{\text{mem}}^{-1}$) is always better (no intermediate fitness maximum). 

\begin{figure}
\begin{centering}
\includegraphics[width=\columnwidth]{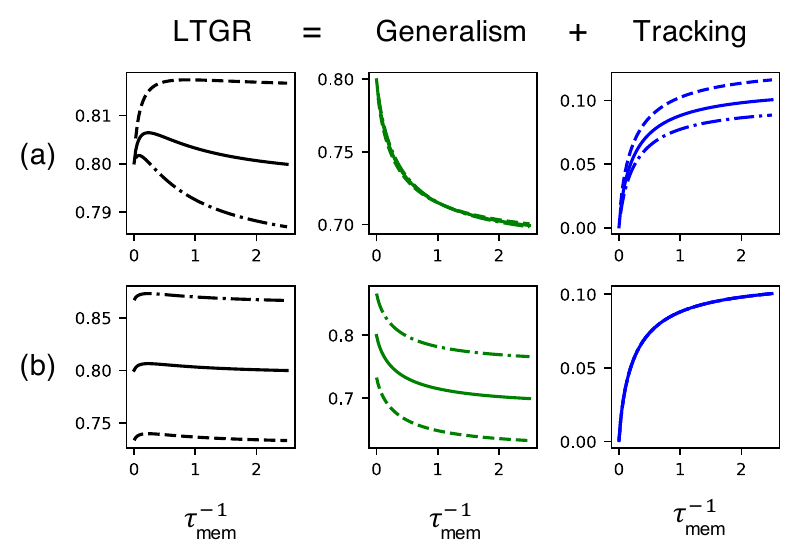}
\par\end{centering}
\caption{{\textbf{Fitness Parsing predicts the parameter dependence of fitness-memory relationship.} \textbf{(a)} Using the parameter set of Fig. \ref{fig: optimal memory}(a) inherited from Ref. \cite{skanata_evolutionary_2016} (solid lines in this figure) as the baseline, we scale up and down the overall environmental switching rates $\theta R_{AB},\theta R_{BA}$: %\textcolor{red}{Use of $\theta$ here before defining it is a bit tough to figure out} : 
$\theta=0.8$ case shown in dash-dotted lines and $\theta=1.2$ case in dash lines. This varies mainly the Tracking gain. \textbf{(b)} We perturb the growth rate under glucose, $g+\phi$: $\phi=-0.2$ shown in dash-dotted lines, and $\phi=+0.2$ in dash lines. This shifts only the Generalism gain. Strengths of parameter perturbations are chosen phenomininological for illustration. %\textcolor{red}{This is a really cool result.}  :)
\label{fig: optimal memory-2}} }
\end{figure}

\textbf{Distinct system controllability:} We give two examples of how the two components can be tuned separately. Scaling the environmental speed (Fig. \ref{fig: optimal memory-2}a) stretches the Tracking curve (via the $\bar{T}$ term) while leaving Generalism approximately invariant. In contrast, shifting the growth baseline (Fig. \ref{fig: optimal memory-2}b) vertically shifts the Generalism contribution but leaves the Tracking term invariant (as the growth difference $\Delta_g$ is fixed). This shows independent controllability of Generalism and Tracking.
%{Our fitness parsing can further predict the effect of parameter perturbations. If the environmental switching rates, $R_{AB}$ and $R_{BA}$, are scaled, our parsing in Eq. \eqref{SFR for LTGR in random environment} indicates that the changes in LTGR is mainly from the tracking gain since the generalism term is affected only implicitly through the long-term average of $f$, which is insensitive to the scaling of environmental switching rates (Fig. \ref{fig: optimal memory-2}a). On the other hand, if one shifts the value of growth rate under glucose $g$, this shifts only the generalism gain since the tracking gain depends only on the growth rate differences $\Delta_g$ (Fig. \ref{fig: optimal memory-2}b).}\\

\section*{Example 4: Orthogonal Control of Adaptation via Environmental Protocols}

So far, we have analyzed how systems optimize internal strategies given environmental statistics. Here, we address the inverse problem: designing environmental protocols to control a fixed adaptive system. We illustrate the utility of our framework using a minimal model of drug resistance.\\

When drugs are applied to killing populations of pathogens, such as bacteria, cancer, or viruses, the resulting selective pressure leads the organism to become drug-resistant.  A sufficiently strong treatment of a drug will drive the rise of a resistant mutant, and subsequent treatment will fail. Accordingly, drug resistance can be viewed as an evolutionary game that we drug-givers are playing against the disease pathogen \cite{stein2023stackelberg}. 
Drawing parallels to ``integrated pest management'' \cite{whelan_resistance_2020,gatenby2020integrating}, we use a toy model, similar to the one in \cite{robertson-tessi_feasibility_2023}, to demonstrate how evolutionary strategies can be explicitly designed to suppress resistance.\\
%An important successful example is ``integrated pest management'' \cite{whelan_resistance_2020,gatenby2020integrating}, where the problem of pesticide resistance has been controlled through better strategies explicitly designed to fight resistance. We now show how a toy model of drug resistance, similar to the one in \cite{robertson-tessi_feasibility_2023}, can be decomposed into components, to ultimately improve evolution-inspired treatment schedules. \\

Consider a pathogen with drug-sensitive population $N_S$ and drug-resistant population $N_R$. We assume that both subpopulations grow at rate $g$ without drug and that the sensitive one reduces to $g-\delta$ when the drug is on. We also assume that that switching from sensitive to resistant only happens under drug, with rate $r$, and re-sensitization happens only when there's no drug, with rate $s$. For a large population, the dynamics can be described by ODEs: 
\begin{subequations}
    \begin{align}
        \dot{N}_S &= (g-\delta D)~N_S -r DN_S + s (1-D)N_R\\
        \dot{N}_R &= g ~N_R +r DN_S - s (1-D)N_R
    \end{align}
\end{subequations}where $D=1$ or $0$ represents drug on or off. 
%As before, we then rewrite them in terms of the total population $N$ and the population fraction of drug-resistant cells $f$ for drug-on, \begin{subequations}
%    \begin{align}
%    \dot{N}&=[g-(1-f)~\delta] N;\\
%    \dot{f}&= f (1-f)~\delta+(1-f)~r,
%\end{align}
%\end{subequations}
%and drug-off, 
%    \begin{equation}
%    \dot{N}=g~ N \text{ and }
%    \dot{f}=- s ~f.
%\end{equation}
Denoting $\pi_{\text{on}}$ as the time fraction of drug-on, we parse the population long-term growth rate under Markov stochastic or periodic drug on/off switching:
\begin{equation}
    \text{LTGR} = [g-\langle 1-f \rangle~\pi_{\text{on}}~ \delta]+(\text{Tracking}) \label{eq: LTGR eq for the killing drug}
\end{equation}
Depending on the protocol, the Tracking term takes the form $\pi_{\text{on}} \pi_{\text{off}} ~(\bar{T}\cdot \bar{J}) ~\delta$ (Markov random) or $\pi_{\text{on}} \pi_{\text{off}} ~(\mathcal{K}_{\text{env}}\circ \bar{J}) ~\delta$ (periodic) \cite{SI}. Crucially, regardless of the protocol, the Tracking gain scales linearly with the environmental switching period. As in Example 3, this Tracking is driven by both differential growth ($g$ versus $g-\delta$) and responsive switching ($r$ versus $s$).\\

\begin{figure}
\begin{centering}
\includegraphics[width= \columnwidth]{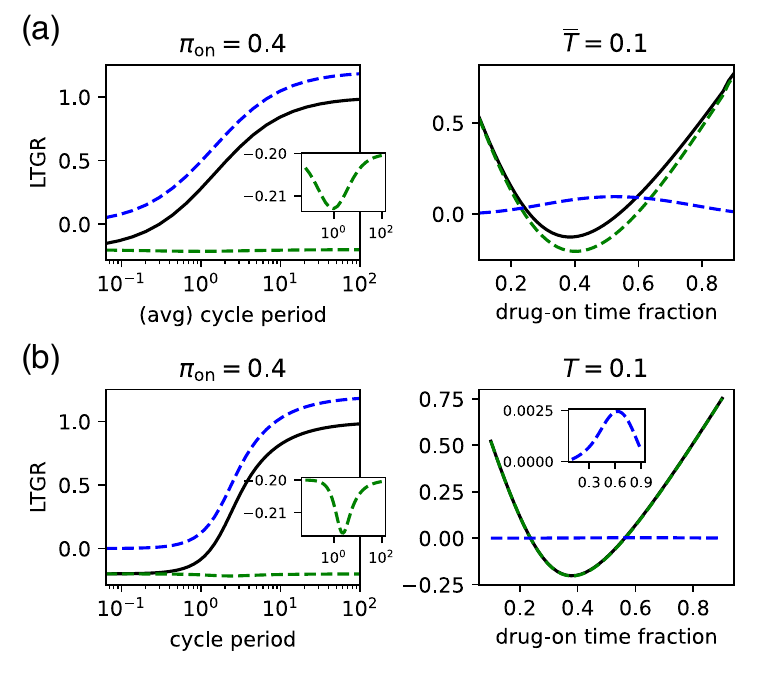}
\par\end{centering}
\caption{\textbf{Fast environment switching reduces Tracking gain whereas optimal environment time fraction reduces Generalism gain.} \textbf{(a)} shows the total long-term population growth rate of drug sensitive or resistant cells under Markov random on/off drug switching. On the left, we vary the average cycle period $\bar{T}=\bar{T}_{\text{on}} + \bar{T}_{\text{off}}$ where $\bar{T}_{\text{on/off}}$ represents long-term average drug on/off intervals and fix the drug-on time fraction $\pi_{\text{on}}=\bar{T}_{\text{on}}/\bar{T}$. On the right, we vary $\pi_{\text{on}}$ and fix $\bar{T}$.   \textbf{(b)} is similar to (a) but under periodic environment: $T=T_{\text{on}}+T_{\text{off}}$ where $T_{\text{on/off}}$ is the fixed interval of on/off and $\pi_{\text{on}}=T_{\text{on}}/T$. {Blue dashed lines are the Tracking components whereas the green dashed lines are the Generalism components.} Parameters in both (a) and (b) are $g=1,\delta=5,r=1,s=3$. See Appendix C3 of SM \cite{SI} for how these are chosen. The time unit is chosen to be 100 days, i.e. rate quantities are per 100 days. 
\label{fig: drug resistance} }
\end{figure}

\textbf{Distinct environmental controllability:}
This orthogonal controllability holds across different protocols (Fig. \ref{fig: drug resistance}).
\textbf{First, rapid switching minimizes Tracking.} As the cycle period vanishes ($\bar{T} \to 0$), the Tracking gain diminishes, leaving growth dominated by the Generalism baseline. This aligns with the ``alternating therapy'' strategy in cancer \cite{goldie1982rationale,strobl2023treatment,perry2012perry}, where rapid alternation of non-cross-resistant drugs is often recommended. While clinical outcomes have been mixed, showing both successes \cite{glick1998mopp,grier2003addition} and failures \cite{bonadonna1995sequential,aisner1995combination,de1995cmf}, our framework offers a physical interpretation: efficacy hinges on minimizing the Tracking flux $J$ driven by re-sensitization. We expect that extending this analysis---from our minimal two-state model to realistic multi-state systems---will help reconcile these disparate findings, predicting optimal protocols by identifying specific regimes where Tracking is effectively suppressed.\\

\textbf{Second, an optimal time fraction minimizes Generalism.}
Equation \eqref{eq: LTGR eq for the killing drug} shows that the Generalism component is minimized at an intermediate drug-on fraction $\pi_{\text{on}}$. In the limits ($\pi_{\text{on}} \to 0$ or $1$), the population adapts to a static environment (growth rate $g$). At intermediate fractions, however, the population retains a sub-population of sensitive cells ($\langle 1-f \rangle > 0$) while still facing frequent drug pressure ($\pi_{\text{on}} > 0$). This combination maximizes the ``cost'' term $\langle 1-f\rangle \pi_{\text{on}} \delta$, effectively suppressing the Generalism baseline.\\

This analysis suggests a rational, two-pronged strategy for suppressing resistance: use \textbf{fast switching} to eliminate the Tracking axis, and an \textbf{optimal duty cycle} ($\pi_{\text{on}}$) to minimize fitness along the Generalism axis.

\section*{Discussion}

\textbf{Quantifying Adaptivity via Observable Coordinates.} Our framework provides a practical principle for experimental quantification: even without measuring microscopic nonequilibrium fluxes, the Tracking contribution can be computed as the residual fitness:
\begin{equation}
\text{Tracking} = \text{LTGR}_{\text{measured}} - \text{Generalism}(\pi_E, \pi_x).
\end{equation}
By subtracting the calculable baseline of Generalism from the measured total growth rate, one can isolate the specific contribution of Tracking, providing a simple yet powerful metric to gauge the ``nonequilibrium advantage'' of systems adapting the varying environments.\\

\textbf{Experimental Connections and Robustness.}
The inputs required for our analysis---such as growth rates and switching rates in the model---are standard variables that can be measured or fitted from time-series data in microfluidic and synthetic setups \cite{nevozhay_mapping_2012,lambert_memory_2014}. A key strength of this framework is its robustness: while our examples explored wide parameter ranges to illustrate distinct regimes (e.g., from slow to fast switching), the predicted qualitative trends---specifically the distinct sensitivities of Tracking to time-scales and Generalism to biases---rely on the generic structure of the nonequilibrium dynamics, not on fine-tuned parameter values. This suggests that the orthogonal control strategies we propose are generic across different adaptive systems.\\

\textbf{Cyclic Origins of Tracking.}
The Tracking component is sustained by nonequilibrium fluxes ($J$). While connecting these fluxes to precise metabolic costs requires specific energetic models, decomposing fluxes into cycles allows us to further dissect the distinct mechanisms of Tracking. As detailed in Appendix C1 \cite{SI}, the cycle decomposition of fluxes \cite{jiang_mathematical_2004} partitions Tracking gain based on whether the driving cycles are intrinsic to the system or involve system-environment coupling. The former represents the system's ``intrinsic nonequilibriumness,'' while the latter represents the coupling. Appendix C1 shows that Tracking remains possible even when all intrinsic cycle affinities are zero \cite{SI}, demonstrating that the nonequilibrium Tracking can be generated solely by system-environment coupling.\\

\textbf{Conclusion.} By mapping fitness onto the orthogonal, measurable coordinates of Generalism and Tracking, this work constructs the navigational map for the optimization and control of adaptation, determining when organisms---or engineered systems---should invest in static robustness versus dynamic responsiveness.\\

\section*{Acknowledgments}
We thank  G\'abor Bal\'{a}zsi {and the anonymous referees} for insightful feedback and Corey Weistuch for pointing us to Ref. \cite{robertson-tessi_feasibility_2023}, which led us to develop the model in Example 4. We are grateful for the financial support from the Laufer Center for Physical and Quantitative Biology at Stony Brook, the John Templeton Foundation  (Grant ID 62564), and NIH (Grant RM1-GM135136).
%{(Ken, any grant that we need to mention here?)} \textcolor{red}{Please check with Doug.  I think support for Charles was from Templeton, and for Ying-Jen was from the Laufer Center, but he'll have the details.}

\section*{Data Availability Statement}
The codes producing the plotted data can be found at \cite{Codes}.

\section*{Authors Contribution}
All authors conceptualized the project together. YJY developed the theory and derived the main results. CK validated and refined the results and developed the model used in Example 4. KD supervised all developments. All authors contributed equally to writing and editing the manuscript.

%\bibliography{main-submitted.bbl}
\bibliography{NEQ-Adaptation}

\end{document}